 \documentclass[final,1p,times]{elsarticle}

\usepackage{amsmath}
\usepackage{amssymb}
\usepackage{amsthm}
\usepackage{amsbsy}
\usepackage{threeparttable}
\usepackage{algorithmicx,algorithm}
\usepackage{algpseudocode}

\newtheorem{sec2_thm1}{Theorem} [section]
\newtheorem{sec2_thm2}[sec2_thm1]{Theorem}
\newtheorem{sec2_thm3}[sec2_thm1]{Theorem}
\newtheorem{sec2_thm4}[sec2_thm1]{Theorem}

\newtheorem{sec3_cor1x}{Corollary} [section]
\newtheorem{sec3_cor2x}[sec3_cor1x]{Corollary}
\newtheorem{sec3_cor1}[sec3_cor1x]{Corollary}

\newtheorem{sec3_lemma1x}{Lemma} [section]
\newtheorem{sec3_lemma2x}[sec3_lemma1x]{Lemma}
\newtheorem{sec3_lemma1}[sec3_lemma1x]{Lemma}
\newtheorem{sec3_lemma2}[sec3_lemma1x]{Lemma}
\newtheorem{sec3_lemma3}[sec3_lemma1x]{Lemma}
\newtheorem{sec3_lemma4}[sec3_lemma1x]{Lemma}
\newtheorem{sec3_lemma5}[sec3_lemma1x]{Lemma}
\newtheorem{sec3_lemma6}[sec3_lemma1x]{Lemma}

\newtheorem{sec3_lemma12_a0}[sec3_lemma1x]{Lemma}
\newtheorem{sec3_lemma12}[sec3_lemma1x]{Lemma}

\begin{document}

\begin{frontmatter}



\title{On the complete weight enumerators of some linear codes with a few weights}


\author[whut]{Minglong ~Qi\corref{cor1}}
\ead{mlqiecully@163.com}
\author[whut]{Shengwu ~Xiong}
\author[whut]{Jingling ~Yuan}
\author[whut]{Wenbi ~Rao}
\author[whut]{Luo ~Zhong}
\cortext[cor1]{Corresponding Author}
\address[whut]{School of Computer Science and Technology,  Wuhan University of Technology, Mafangshan West Campus, 430070 Wuhan City, China}

\begin{abstract}
Linear codes with a few weights have important applications in authentication codes, secret sharing, consumer electronics, etc.. The determination of the parameters such as Hamming weight distributions and complete weight enumerators of linear codes are  important research topics. 
In this paper, we consider  some classes of linear codes with a few weights and determine the complete weight enumerators from which the corresponding Hamming weight distributions are derived with help of some sums involving  Legendre symbol. 
\end{abstract}


\end{frontmatter}


\section{Introduction}\label{intro}
Throughout this paper, let $ p $ be an odd prime and $ q=p^{e} $ with $ e=2m $, where  $ e $ and $ m $ are two positive integers. Two symbols, $ \alpha $ and $ d $, are kept unchanged as well where $  \alpha $ is a positive integer and $ d=\gcd(\alpha,e)$. An $ [n,k,s] $ linear code $ \mathcal{C} $ over $ \mathbb{F}_{p} $ is a $ k $-dimensional subspace of vector space $ \mathbb{F}_{p}^{n} $ with  minimum Hamming distance $ s $. Let $ A_{i} $ denote the number of codewords with Hamming weight $ i $ in $ \mathcal{C} $, then the polynomial $1+A_{1}z+ A_{2}z^{2}+\cdots+  A_{n}z^{n}$ is referred to as the \textit{weight enumerator} of $ \mathcal{C} $, and the sequence $\lbrace A_{1},\cdots,A_{n} \rbrace$ the \textit{weight distribution} of $ \mathcal{C} $. The code $ \mathcal{C} $ is said to be $ t $-\textit{weight} if there are $ t $ non-zero components in its weight distribution. 

Let $ \mathbf{c}=(c_{0},c_{1},\cdots,c_{n-1})$ be a codeword of the linear code $\mathcal{C}  $. Then, the complete weight enumerator of $ \mathbf{c} $ is a monomial polynomial in the unknowns $ w_{i} (0\leq i\leq p-1)$ given below
\begin{equation}\label{intro_cwe_def0}
w(\mathbf{c})=w_{0}^{t_{0}}w_{1}^{t_{1}}\cdots w_{p-1}^{t_{p-1}}
\end{equation}
where $ t_{i}(0\leq i\leq p-1) $ denotes the number of components of $ \mathbf{c} $ that are equal to $ i $. From the complete weight enumerator of $ \mathbf{c} $ can be defined the complete weight enumerator of the linear code $\mathcal{C}  $ as a homogeneous polynomial of degree $ n $ (see \cite{Bib18,Bib19}):
\begin{equation}\label{intro_cwe_def1}
\mathit{CWE}(\mathcal{C})=\sum_{\mathbf{c}\in\mathcal{C}}w(\mathbf{c}).
\end{equation}
Note that  $ \mathcal{C} $ can be partitioned into disjoint union of some subsets such that every codeword of a such subset has the same complete weight enumerator. Hence, (\ref{intro_cwe_def1}) may be transformed into the following equation:
\begin{equation}\label{intro_cwe_def2}
\mathit{CWE}(\mathcal{C})=\sum\limits_{\mathbf{t}\in \mathbb{F}_{p}^{p}\setminus \lbrace \mathbf{0}\rbrace }\mathfrak{F}(t_{0},t_{1},\cdots,t_{p-1})w_{0}^{t_{0}}w_{1}^{t_{1}}\cdots w_{p-1}^{t_{p-1}}
\end{equation}
where $ \mathbf{t}=(t_{0},t_{1},\cdots,t_{p-1})$, $ \mathbf{0} $ is the zero vector of $ \mathbb{F}_{p}^{p} $, and $ \mathfrak{F}(t_{0},t_{1},\cdots,t_{p-1}) $ denotes the number of the codewords that have $ w_{0}^{t_{0}}w_{1}^{t_{1}}\cdots w_{p-1}^{t_{p-1}}  $ as their complete weight enumerator. The composition of the codeword $  \mathbf{c}$ associated with its complete weight enumerator given in (\ref{intro_cwe_def0}) is a $ p- $tuple,  comp$(\mathbf{c}) =(t_{0},t_{1},\cdots,t_{p-1})$ \cite{Bib18}. $ \mathfrak{F}(t_{0},t_{1},\cdots,t_{p-1}) $ is referred to as \textit{the frequency of the composition} $ (t_{0},t_{1},\cdots,t_{p-1}) $, or as \textit{the frequency of the complete weight enumerator} $ w_{0}^{t_{0}}w_{1}^{t_{1}}\cdots w_{p-1}^{t_{p-1}} $, and $ t_{i}(0\leq i\leq p-1) $ is referred to as \textit{the exponent of symbol} $ i $. The generic codeword $ \mathbf{c} $ may have different composition depending on what constraint is applied to it. Therefore, the item $ \mathfrak{F}(t_{0},t_{1},\cdots,t_{p-1})w_{0}^{t_{0}}w_{1}^{t_{1}}\cdots w_{p-1}^{t_{p-1}} $ defined in (\ref{intro_cwe_def2}), is sometimes called \textit{the contribution of the case} indicating that constraint.

The complete weight enumerators are the important parameters in coding theory from which the corresponding Hamming weight enumerators can be  deduced. While studying the Reed-Solomon codes, Blake and Kith showed that the complete weight enumerators could be an useful tool for soft decision decoding\cite{B20}. In \cite{B21}, Helleseth and Kholosha demonstrated that the study of monomial and quadratic bent functions was related to the determination of complete weight enumerators of linear codes. Ding et al. \cite{B22,B23} showed that the complete weight enumerators could be used to compute the deception probabilities of certain authentication codes.  In \cite{B24,B25}, Kuzmin and Nechaev investigated the generalized Kerdoc codes and related linear codes over Galois rings and determined their complete weight enumerators. In \cite{B26,B27,B28}, the complete weight enumerators of some constant composition codes were studied and determined.

Linear codes with a few weights are of important in secret sharing \cite{B29,B30}, authentication codes \cite{B22}, association schemes \cite{Bib2} and strongly regular graphs \cite{B31}. In \cite{Bib3,Bib4}, Ding et al. introduced a general construction method called the defining set method,  by which the  authors of \cite{Bib5, Bib6, Bib7,Bib8} could construct some new linear codes with at most 3-weights, determined their weight enumerators, and studied their application in the information theory. Very recently, in \cite{Bib8,B32,B33,B34,B35,B36}, the authors constructed some new linear codes with a few weights and settled their complete weight enumerators by using the defining set method.

Let $ D $ be an $ n $-subset of $ \mathbb{F}_{q} $. A linear code of length $ n $ over $ \mathbb{F}_{p} $ can be constructed by the following definition 
\begin{equation*}
\mathcal{C}_{D}=\bigl\lbrace\bigl(Tr(xd_{1}),Tr(xd_{2}),\cdots,Tr(xd_{n})\bigr):x\in \mathbb{F}_{q} \bigr\rbrace,
\end{equation*}
where $ Tr $ denotes the trace function from $ \mathbb{F}_{q} $ to $ \mathbb{F}_{p} $, and $ d_{i}\in D (1\leq i\leq n) $. The set $ D $ is called \textit{the defining set} of the code $ \mathcal{C}_{D} $. 

Let $ f(X)=a^{p^{\alpha}}X^{p^{2\alpha}}+aX $ be a mapping from  $ \mathbb{F}_{q} $ to  $ \mathbb{F}_{q} $, where $ a\in   \mathbb{F}_{q}^{*}  $. 
Coulter \cite{Bib9,Bib10} obtained some important results on exponential sums over  $ \mathbb{F}_{q} $ related to $ f(X) $.
 Based on the results of Coulter  \cite{Bib9,Bib10} and  \textit{the defining set method}, Q. Wang et al constructed two classes of linear codes with a few weights and determined the complete weight enumerators and the weight distributions \cite{Bib8}.  The defining set in \cite{Bib8} is below
\begin{equation}\label{wang_definig_set}
D=\biggl\lbrace x\in \mathbb{F}_{q} : Tr(x^{p^{\alpha}+1})=a \biggr\rbrace,
\end{equation}
and the corresponding linear codes are defined as
\begin{equation}\label{wang_LD}
\mathcal{C}_{D}=\biggl\lbrace\bigl(Tr(xd_{1})+c,Tr(xd_{2})+c,\cdots,Tr(xd_{n})+c\bigr):x\in \mathbb{F}_{q} \biggr\rbrace,
\end{equation}
where $ a,c\in \mathbb{F}_{p} $, and $ d_{i}\in D  (1\leq i\leq n) $.

 \section{Main Theorems}

 \begin{sec2_thm1}\label{s2_thm1}
 In (\ref{wang_definig_set}) and (\ref{wang_LD}), let $ a=0 $ and $ c\ne 0 $. In addition, suppose that $ \frac{m}{d} $ is  odd. Then, the code $ \mathcal{C}_{D} $ of (\ref{wang_LD}) is a $ [p^{e-1}-(p-1)p^{m-1}-1,e+1] $ linear code whose weight distribution is listed in Table \ref{tab_s2_thm1}, and whose complete weight enumerator is 
 \begin{equation}\label{cwe_a0c_mdodd}
 \begin{split}
 \mathit{CWE}(\mathcal{C}_{D})&=w_{c}^{p^{e-1}-(p-1)p^{m-1}-1}+\bigl(p^{e-1}-(p-1)p^{m-1}-1\bigr)w_{c}^{p^{e-2}-(p-1)p^{m-1}-1}\prod_{\substack{0\le i\leq p-1\\i \ne c}}w_{i}^{p^{e-2}}\\
 &+(p-1)(p^{e-1}+p^{m-1})w_{c}^{p^{e-2}-1}\prod_{\substack{0\le i\leq p-1\\i \ne c}}w_{i}^{p^{e-2}-p^{m-1}}.
 \end{split}
 \end{equation}
 \end{sec2_thm1}
  \begin{table}[tbh]
   \caption{Weight distribution of the code of  Theorem \ref{s2_thm1}}
   \label{tab_s2_thm1}
   {\renewcommand{\tabcolsep}{0.15cm}
   \begin{center}
   \begin{tabular}{|c|c|}
   \hline
   weight $ w $ & multiplicity $ A_{w} $ \\
   \hline
   $ (p-1)(p^{e-2}-p^{m-1})-1 $ & $ p^{e-1}-(p-1)p^{m-1}-1 $\\
   \hline
   $ (p-1)p^{e-2}-(p-2)p^{m-1}-1 $ & $ (p-1)(p^{e-1}+p^{m-1}) $\\
   \hline
  $ p^{e-1}-(p-1)p^{m-1}-1 $ & $ 1 $\\
   \hline
   \end{tabular}
   \end{center}
   }
   \end{table}

    \begin{sec2_thm2}\label{s2_thm2}
    In (\ref{wang_definig_set}) and (\ref{wang_LD}), let $ a\ne 0 $ and $ c\ne 0 $. In addition, suppose that $ \frac{m}{d} $ is odd. Then, the code $ \mathcal{C}_{D} $ of (\ref{wang_LD}) is a $ [p^{e-1}+p^{m-1},e+1] $ linear code whose weight distribution is listed in Table \ref{tab_s2_thm2},  and whose complete weight enumerator is 
     \begin{equation}\label{cwe_ac_mdodd}
     \begin{split}
     \mathit{CWE}(\mathcal{C}_{D})&=w_{c}^{p^{e-1}+p^{m-1}}+
     \bigl(p^{e-1}-(p-1)p^{m-1}-1\bigr)w_{c}^{p^{e-2}+p^{m-1}}\prod_{\substack{1\leq i\leq p-1}}w_{i+c}^{p^{e-2}}\\
     &+(p^{e-1}+p^{m-1})w_{c}^{p^{e-2}-\bigl(\frac{-1}{p}\bigr)p^{m-1}}\prod_{\substack{1\leq i\leq p-1}}w_{i+c}^{p^{e-2}-\bigl(\frac{i^{2}-c^{2}}{p}\bigr)p^{m-1}}\\
     &+(p^{e-1}+p^{m-1})\sum_{\substack{1\leq i\leq p-1\\i\ne (4a)^{-1}c^{2}\\\bigl(\frac{i}{p}\bigr)=\bigl(\frac{a}{p}\bigr)}}w_{c}^{p^{e-2}-\bigl(\frac{-1}{p}\bigr)p^{m-1}}
       \prod_{\substack{1\leq j\leq p-1}}w_{j+c}^{p^{e-2}-\bigl(\frac{j^{2}-4ai}{p}\bigr)p^{m-1}}\\
       &+(p^{e-1}+p^{m-1})\sum_{\substack{1\leq i\leq p-1\\\bigl(\frac{i}{p}\bigr)\ne \bigl(\frac{a}{p}\bigr)}} w_{c}^{p^{e-2}+\bigl(\frac{-1}{p}\bigr)p^{m-1}}
        \prod_{\substack{1\leq j\leq p-1}}w_{j+c}^{p^{e-2}-\bigl(\frac{j^{2}-4ai}{p}\bigr)p^{m-1}},
     \end{split}
     \end{equation}
     where the index of each unknown $ w $ is reduced modulo $ p $, and $ \bigl(\frac{*}{p}\bigr) $ denotes the Legendre symbol modulo $ p $.
    \end{sec2_thm2}
    \begin{table}[tbh]
       \caption{Weight distribution of the code of  Theorem \ref{s2_thm2}}
       \label{tab_s2_thm2}
       \begin{center}
       \begin{tabular}{|c|c|}
       \hline
       weight $ w $ & multiplicity $ A_{w} $ \\
        \hline
       $ (p-1)p^{e-2} $ & $ \frac{1}{2}(p-1)(p^{e-1}+p^{m-1}) $\\
       \hline
       $ (p-1)p^{e-2}+p^{m-1} $ & $ 2p^{e-1}-(p-2)p^{m-1}-1 $\\
       \hline
       $ (p-1)p^{e-2}+2p^{m-1} $ & $ \frac{1}{2}(p-3)(p^{e-1}+p^{m-1}) $\\
       \hline
       $ p^{e-1}+p^{m-1} $ & $ 1 $\\
        \hline
       \end{tabular}
       \end{center}
       \end{table}

  \begin{sec2_thm3}\label{s2_thm3}
   In (\ref{wang_definig_set}) and (\ref{wang_LD}), let $ a= 0 $ and $ c\ne 0 $. In addition, suppose that $ \frac{m}{d} $ is  even. Then, the code $ \mathcal{C}_{D} $ of (\ref{wang_LD}) is a $ [p^{e-1}-(p-1)p^{m+d-1}-1,e+1] $ linear code whose weight distribution is listed in Table \ref{tab_s2_thm3}, and whose complete weight enumerator is
    \begin{equation}\label{cwe_a0c_mdeven}
    \begin{split}
    \mathit{CWE}(\mathcal{C}_{D})&=w_{c}^{p^{e-1}-(p-1)p^{m+d-1}-1}+(p^{e}-p^{e-2d})w_{c}^{p^{e-2}-(p-1)p^{m+d-2}-1}
    \prod_{\substack{0\leq i\leq p-1\\i\ne c}}w_{i}^{p^{e-2}-(p-1)p^{m+d-2}}\\
    &+\bigl(p^{e-2d-1}-(p-1)p^{m-d-1}-1\bigr)w_{c}^{p^{e-2}-(p-1)p^{m+d-1}-1}
        \prod_{\substack{0\leq i\leq p-1\\i\ne c}}w_{i}^{p^{e-2}}\\
    &+(p-1)\bigl(p^{e-2d-1}+p^{m-d-1}\bigr)w_{c}^{p^{e-2}-1}\prod_{\substack{0\leq i\leq p-1\\i\ne c }}w_{i}^{p^{e-2}-p^{m+d-1}}.
    \end{split}
    \end{equation}
  \end{sec2_thm3}
  \begin{table}[tbh]
  \caption{Weight distribution of the code of  Theorem \ref{s2_thm3}}
         \label{tab_s2_thm3}
         \begin{center}
         \begin{tabular}{|c|c|}
         \hline
         weight $ w $ & multiplicity $ A_{w} $ \\
          \hline
         $ (p-1)(p^{e-2}-(p-1)p^{m+d-2})-1 $ & $p^{e}-p^{e-2d} $\\
         \hline
         $ (p-1)(p^{e-2}-p^{m+d-1})-1 $ & $ p^{e-2d-1}-(p-1)p^{m-d-1}-1 $\\
         \hline
         $ (p-1)p^{e-2}-(p-2)p^{m+d-1}-1 $ & $(p-1)(p^{e-2d-1}+p^{m-d-1}) $\\
         \hline
         $ p^{e-1}-(p-1)p^{m+d-1}-1 $ & $ 1 $\\
          \hline
         \end{tabular}
         \end{center}
   \end{table}

   \begin{table}[h]
       \caption{Weight distribution of the code of  Theorem \ref{s2_thm4}}
              \label{tab_s2_thm4}
              \begin{center}
              \begin{tabular}{|c|c|}
              \hline
              weight $ w $ & multiplicity $ A_{w} $ \\
               \hline
              $ (p-1)(p^{e-2}+p^{m+d-2}) $ & $p^{e}-p^{e-2d} $\\
              \hline
              $ (p-1)p^{e-2}+p^{m+d-1} $ & $ 2p^{e-2d-1}-(p-2)p^{m-d-1}-1 $\\
              \hline
              $ (p-1)p^{e-2}+2p^{m+d-1} $ & $\frac{1}{2}(p-3)(p^{e-2d-1}+p^{m-d-1}) $\\
              \hline
              $ (p-1)p^{e-2} $ & $ \frac{1}{2}(p-1)(p^{e-2d-1}+p^{m-d-1}) $\\
              \hline
              $ p^{e-1}+p^{m+d-1} $ & $ 1 $\\
               \hline
              \end{tabular}
              \end{center}
        \end{table}
        
   \begin{sec2_thm4}\label{s2_thm4}
     In (\ref{wang_definig_set}) and (\ref{wang_LD}), let $ (a,c)\in \mathbb{F}_{p}^{*}\times \mathbb{F}_{p}^{*} $. In addition, suppose that $ \frac{m}{d} $ is even. Then, the code $ \mathcal{C}_{D} $ of (\ref{wang_LD}) is a $ [p^{e-1}+p^{m+d-1},e+1] $ linear code whose weight distribution is listed in Table \ref{tab_s2_thm4}, and whose complete weight enumerator is 
     \begin{equation}\label{Fac5Even}
       \begin{split}
       \mathit{CWE}(\mathcal{C}_{D})&=w_{c}^{p^{e-1}+p^{m+d-1}}+(p^{e}-p^{e-2d})\prod_{\substack{0\leq i\leq p-1}}w_{i}^{p^{e-2}+p^{m+d-2}}\\
       &+\bigl(p^{e-2d-1}-(p-1)p^{m-d-1}-1\bigr)w_{c}^{p^{e-2}+p^{m+d-1}}\prod_{\substack{1\leq i\leq p-1}}w_{i+c}^{p^{e-2}}\\
       &+(p^{e-2d-1}+p^{m-d-1})w_{c}^{p^{e-2}-\bigl(\frac{-1}{p}\bigr)p^{m+d-1}}\prod_{\substack{1\leq i\leq p-1}}w_{i+c}^{p^{e-2}-\bigl(\frac{i^{2}-c^{2}}{p}\bigr)p^{m+d-1}}\\
       &+(p^{e-2d-1}+p^{m-d-1})\sum_{\substack{1\leq i\leq p-1\\i\ne (4a)^{-1}c^{2}\\\bigl(\frac{i}{p}\bigr)=\bigl(\frac{a}{p}\bigr)}}w_{c}^{p^{e-2}-\bigl(\frac{-1}{p}\bigr)p^{m+d-1}}
       \prod_{\substack{1\leq j\leq p-1}}w_{j+c}^{p^{e-2}-\bigl(\frac{j^{2}-4ai}{p}\bigr)p^{m+d-1}}\\
       &+(p^{e-2d-1}+p^{m-d-1})\sum_{\substack{1\leq i\leq p-1\\\bigl(\frac{i}{p}\bigr)\ne \bigl(\frac{a}{p}\bigr)}} w_{c}^{p^{e-2}+(\frac{-1}{p})p^{m+d-1}}
        \prod_{\substack{1\leq j\leq p-1}}w_{j+c}^{p^{e-2}-\bigl(\frac{j^{2}-4ai}{p}\bigr)p^{m+d-1}},
        \end{split}
       \end{equation}
     where the index of each unknown $ w $ is reduced modulo $ p $, and $ \bigl(\frac{*}{p}\bigr) $ denotes the Legendre symbol modulo $ p $.
    \end{sec2_thm4}

   \section{Proof of the main theorems}
   
   This section is devoted to prove the main theorems of the previous section, before that some concepts and basic results on group character and a series of needed lemmas, such as the work of Coulter on exponential sum over finite field \cite{Bib9,Bib10}, some intermediate results in \cite{Bib7}, will be presented. At first, we introduce some sums of Legendre's Symbols needed in the estimation of the Hamming weights from the corresponding complete weight enumerators.
   
   \begin{sec3_lemma1x}[\cite{Bib17}]\label{sec3_legendre_symbol}
   Let $ p $ be an odd prime, and $ a,b,c $ be any integers. Then
   \begin{equation*}
   \sum_{x=0}^{p-1}\biggl(\dfrac{ax^{2}+bx+c}{p}\biggr)=
   \begin{cases}
   -\biggl(\dfrac{a}{p}\biggr)\quad &\text{if}\ p\nmid b^{2}-4ac\\
    (p-1)\biggl(\dfrac{a}{p}\biggr)\quad &\text{if}\ p\mid b^{2}-4ac
   \end{cases}
   \end{equation*}
   \end{sec3_lemma1x}
   where $ \biggl(\dfrac{*}{p}\biggr) $ denotes Legendre Symbol modulo $ p $.
   
   Let $ p $ be an odd prime, and $ a,c,i\in \mathbb{F}_{p}^{*} $. Define the following two sums:
   \begin{equation*}
   S_{p}(a,i)=\sum_{j=1}^{p-1}\biggl(\dfrac{j^{2}-4ai}{p}\biggr),
   L_{p}(c)=\sum_{j=1}^{p-1}\biggl(\dfrac{j^{2}-c^{2}}{p}\biggr).
   \end{equation*}
   
   Then, we can obtain the next two corollaries from Lemma \ref{sec3_legendre_symbol} whose proofs are omitted:
   \begin{sec3_cor1x}\label{sec3_Legendre_symbol_spai}
   \begin{enumerate}[(1)]
   \item $ p\equiv 1\pmod 4 $.
    \begin{equation*}
      S_{p}(a,i)=
      \begin{cases}
      -2\quad &\text{if}\ \biggl(\dfrac{a}{p}\biggr)=\biggl(\dfrac{i}{p}\biggr)\\
      0\quad &\text{if}\ \biggl(\dfrac{a}{p}\biggr)\ne\biggl(\dfrac{i}{p}\biggr)
      \end{cases}
   \end{equation*}
   \item $ p\equiv 3\pmod 4 $.
       \begin{equation*}
         S_{p}(a,i)=
         \begin{cases}
         0\quad &\text{if}\ \biggl(\dfrac{a}{p}\biggr)=\biggl(\dfrac{i}{p}\biggr)\\
         -2\quad &\text{if}\ \biggl(\dfrac{a}{p}\biggr)\ne\biggl(\dfrac{i}{p}\biggr)
         \end{cases}
      \end{equation*}
   \end{enumerate}
   \end{sec3_cor1x}
  
  \begin{sec3_cor2x}\label{sec3_Legendre_symbol_Lpc}
  \begin{equation*}
        L_{p}(c)=
        \begin{cases}
        -2\quad &\text{if}\ p\equiv 1\pmod 4\\
        0\quad &\text{if}\ p\equiv 3\pmod 4
        \end{cases}
     \end{equation*}
  \end{sec3_cor2x}
  
  We need also the cardinalities of the following four sets in the derivation of the Hamming weights from their complete weight enumerators:
  \begin{align*}
  N_{p}^{+}(a,c)&=\biggl\lbrace 1\leq i\leq p-1:\biggl(\dfrac{c^{2}-4ai}{p}\biggr)=1,i\ne \dfrac{c^{2}}{4a},\biggl(\dfrac{i}{p}\biggr)=\biggl(\dfrac{a}{p}\biggr)\biggr\rbrace,\\
  N_{p}^{-}(a,c)&=\biggl\lbrace 1\leq i\leq p-1:\biggl(\dfrac{c^{2}-4ai}{p}\biggr)=-1,i\ne \dfrac{c^{2}}{4a},\biggl(\dfrac{i}{p}\biggr)=\biggl(\dfrac{a}{p}\biggr)\biggr\rbrace,\\
  M_{p}^{+}(a,c)&=\biggl\lbrace 1\leq i\leq p-1:\biggl(\dfrac{c^{2}-4ai}{p}\biggr)=1,i\ne \dfrac{c^{2}}{4a},\biggl(\dfrac{i}{p}\biggr)\ne\biggl(\dfrac{a}{p}\biggr)\biggr\rbrace,\\
    M_{p}^{-}(a,c)&=\biggl\lbrace 1\leq i\leq p-1:\biggl(\dfrac{c^{2}-4ai}{p}\biggr)=-1,i\ne \dfrac{c^{2}}{4a},\biggl(\dfrac{i}{p}\biggr)\ne\biggl(\dfrac{a}{p}\biggr)\biggr\rbrace,
  \end{align*}
  where $ \biggl(\dfrac{*}{p}\biggr)$ denotes Legendre Symbol, and $ a,c\in \mathbb{F}_{p}^{*} $.
  
  \begin{sec3_lemma2x}[\cite{Bib17}]\label{sec3_npac}
  If $ p\equiv 1\pmod 4 $, then $ \#N_{p}^{+}(a,c)=\frac{p-5}{4} , \#N_{p}^{-}(a,c)=\frac{p-1}{4} $, and $ \#M_{p}^{+}(a,c)=\#M_{p}^{-}(a,c)=\frac{p-1}{4} $, else if $ p\equiv 3\pmod 4 $, then $ \#N_{p}^{+}(a,c)=\#N_{p}^{-}(a,c)=\frac{p-3}{4} , \#M_{p}^{+}(a,c)=\frac{p-3}{4} $ and $ \#M_{p}^{-}(a,c)=\frac{p+1}{4} $. 
  \end{sec3_lemma2x}
   
   An \textit{additive character} of $ \mathbb{F}_{q} $, $ \chi $, is a nonzero function from $ \mathbb{F}_{q} $ to a set of nonzero complex numbers such that for any pair $ (x,y) \in \mathbb{F}_{q}\times \mathbb{F}_{q} $, $ \chi (x+y) =\chi(x)\chi(y)$. In this paper, the complex conjugate of $ \chi $ is denoted by $\overline{\chi} $. For each $ b\in \mathbb{F}_{q}$, an additive character of  $ \mathbb{F}_{q} $ can be define below
   \begin{equation}\label{additive_character}
   \chi_{b}(c)=\varsigma_{p}^{Tr(bc)},\ \text{for all}\ c\in \mathbb{F}_{q},
   \end{equation}
   where $ \varsigma_{p}=e^{\frac{2\pi\sqrt{-1}}{p}} $. In (\ref{additive_character}),  the  character $ \chi_{0} $ is said to be \textit{trivial} since for all $  c\in \mathbb{F}_{q} $, $ \chi_{0}(c) =1$. The character $ \chi_{1} $ is called the \textit{canonical additive character}. It was shown that  any additive character of 
  $ \mathbb{F}_{q} $ can be written as $ \chi_{b}(x)=\chi_{1}(bx) $ \cite[Chapter 5]{Bib15}. In this paper, the canonical additive character is used and its subscript omitted.
  
   The orthogonal property of the additive character over $ \mathbb{F}_{q} $ is resumed in the following \cite[Chapter 5]{Bib15}:
  \begin{equation*}
  \sum_{x\in \mathbb{F}_{q}}\chi_{b}(x)=
  \begin{cases}
  q,&\ \text{if}\ b=0,\\
  0,&\ \text{otherwise.}
  \end{cases}
  \end{equation*}
  
  A \textit{multiplicative character} $ \psi $ of $ \mathbb{F}_{q} $ over $ \mathbb{F}_{q}^{*} $ is a nonzero function from $ \mathbb{F}_{q}^{*} $ to  a set of nonzero complex number such that for any $ (x,y)\in \mathbb{F}_{q}^{*}\times\mathbb{F}_{q}^{*}$,  $ \psi(xy)=\psi(x)\psi(y) $. Let $ \theta $ be a primitive element of $ \mathbb{F}_{q}^{*} $. Then, any multiplicative character over $ \mathbb{F}_{q}^{*} $ can be written as 
  \begin{equation*}
  \psi_{j} (\theta^{k}) =e^{2\pi\sqrt{-1} jk/(q-1)},    
  \end{equation*}
  where $ 0\leq j,k\leq q-2 $. The multiplicative character $ \psi_{(q-1)/2} $ is called \textit{the quadratic character} of $ \mathbb{F}_{q} $, denoted by $ \eta $. In this paper, suppose that $ \eta(0)=0 $. 
  With the canonical additive character and the quadratic character can the quadratic Gauss sum over $ \mathbb{F}_{q} $ be defined by
  \begin{equation}\label{Gauss_sum_def}
  G(\eta,\chi)=\sum_{x\in \mathbb{F}_{q}^{*}}\eta(x)\chi(x).
  \end{equation}
  In order not to confuse with the complex conjugation, let $ \widehat{\chi} $ denote the canonical additive character over $ \mathbb{F}_{p} $, and $ \widehat{\eta} $ the quadratic character over $ \mathbb{F}_{p} $, respectively.
  
  Let $ a\in \mathbb{F}_{q}^{*} $ and $ b\in  \mathbb{F}_{q} $ where $ q=p^{e} $ with $ p $ an odd prime and $ e $ a positive integer. In \cite{Bib10}, Coulter gave the definition of a kind of exponential sum as below
  \begin{equation}\label{Coulter_sum_sab}
  S_{\alpha}(a,b)=\sum_{x\in\mathbb{F}_{q}}\chi(ax^{p^{\alpha}+1}+bx),
  \end{equation}
  and explicit formulae to evaluate that sum which are resume in the next  lemmas:
  \begin{sec3_lemma1}[\cite{Bib10},Theorem 1]\label{s3_lemma1}
  Let $ q $ be odd and suppose $ f(X)=a^{p^{\alpha}}X^{p^{2\alpha}} +aX$ is a permutation polynomial over $ \mathbb{F}_{q} $. Let $ x_{0} $ be the unique solution of the equation $ f(x)=- b^{p^{\alpha}},b\ne 0$. The evaluation of (\ref{Coulter_sum_sab}) partitions into the following two cases.
  \begin{enumerate}[(1)]
  \item If $ e/d  $ is odd then
  \begin{equation*}
   S_{\alpha}(a,b)=
   \begin{cases}
   (-1)^{e-1}\sqrt{q}\eta(-a)\overline{\chi(ax_{0}^{p^{\alpha}+1}})&\qquad\text{if}\ p\equiv 1\pmod 4\\
   (-1)^{e-1}\sqrt{-1}^{3e}\sqrt{q}\eta(-a)\overline{\chi(ax_{0}^{p^{\alpha}+1}})&\qquad\text{if}\ p\equiv 3\pmod 4
    \end{cases}
  \end{equation*}
   \item If $ e/d  $ is even then $ e=2m $, $ a^{(q-1)/(p^{d}+1)}\ne (-1)^{m/d} $ and
     \begin{equation*}
     S_{\alpha}(a,b)=-(-1)^{m/d}p^{m}\overline{\chi(ax_{0}^{p^{\alpha}+1}}).
     \end{equation*}
  \end{enumerate}
  \end{sec3_lemma1} 
  \begin{sec3_lemma2}[\cite{Bib10},Theorem 2]\label{s3_lemma2}
  Let $ q=p^{e} $ be odd and suppose $ f(X)=a^{p^{\alpha}}X^{p^{2\alpha}} +aX$ is not a permutation polynomial over $ \mathbb{F}_{q} $. Then for $ b\ne 0 $ we have $ S_{\alpha}(a,b)=0 $ unless the equation $ f(X)=- b^{p^{\alpha}}$ is solvable. If this equation is solvable, with some solution $ x_{0} $ say, then
  \begin{equation*}
  S_{\alpha}(a,b)=-(-1)^{m/d}p^{m+d}\overline{\chi(ax_{0}^{p^{\alpha}+1}}).
  \end{equation*}
  \end{sec3_lemma2}
  
  Recall that $ d=\gcd(e,\alpha) $, and $ q=p^{e} $. If $ e/d $ is odd or $ e/d $ is even with $ e=2m $ and  $ a^{(q-1)/(p^{d}+1)}\ne (-1)^{m/d} $ , then $ f(X)=a^{p^{\alpha}}X^{p^{2\alpha}} +aX$ is a permutation polynomial over $ \mathbb{F}_{q} $ and $ f(X)=a^{p^{\alpha}}X^{p^{2\alpha}} +aX=0 $ has no nonzero solution over $ \mathbb{F}_{q} $. If $ e/d $ is even with $ e=2m $ and  $ a^{(q-1)/(p^{d}+1)}= (-1)^{m/d} $, then, $ f(X)=a^{p^{\alpha}}X^{p^{2\alpha}} +aX=0 $ has $ p^{2d}-1 $ nonzero solutions in $ \mathbb{F}_{q}^{*} $ \cite[Theorem 4.1]{Bib9}.
  
  \begin{sec3_cor1}\label{s3_cor1}
  Let $ q=p^{e} $ be odd and suppose $ f(X)=X^{p^{2\alpha}} +X$ is not a permutation polynomial over $ \mathbb{F}_{q} $, and that $ f(X)=- b^{p^{\alpha}}$ is solvable for a given $ b\in \mathbb{F}_{q}^{*} $. Denote the set of the solutions as $ \mathfrak{S}_{b} $. Then, for each distinct pair $ (x_{1},x_{2})\in \mathfrak{S}_{b}\times \mathfrak{S}_{b} $, we have $ Tr(x_{1}^{p^{\alpha}+1}) =Tr(x_{2}^{p^{\alpha}+1})=a$, where $ a\in \mathbb{F}_{p} $.
  \end{sec3_cor1}
  \begin{proof}
  From the above discussion, $ f(X)=X^{p^{2\alpha}} +X$ is not a permutation polynomial over $ \mathbb{F}_{q} $ if and only if  $ e/d $ is even with $ e=2m $ and $ m/d $ is even, for that $ a^{(q-1)/(p^{d}+1)}= (-1)^{m/d} $ holds always for $ a=1 $. From  Lemma \ref{s3_lemma2}, it is clear that the value of $ S_{\alpha}(1,b) $ dosn't depend on which solution of $ f(X)=- b^{p^{\alpha}}$ is chosen, so that we have 
  \begin{equation*}
 S_{\alpha}(1,b)=-(-1)^{m/d}p^{m+d}\overline{\chi(x_{1}^{p^{\alpha}+1}})= -(-1)^{m/d}p^{m+d}\overline{\chi(x_{2}^{p^{\alpha}+1}}),
  \end{equation*}
   which implicates  $ Tr(x_{1}^{p^{\alpha}+1}) =Tr(x_{2}^{p^{\alpha}+1}) =a$ for some $ a\in \mathbb{F}_{p} $.
  \end{proof}
  
  We denote the property of Corollary \ref{s3_cor1} as $ Tr(\mathfrak{S}_{b})=a $ which means that for each distinct pair $ (x_{1},x_{2})\in \mathfrak{S}_{b}\times \mathfrak{S}_{b} $, $ Tr(x_{1}^{p^{\alpha}+1}) =Tr(x_{2}^{p^{\alpha}+1})=a$.
  
  When determining the weight distribution, we need to know how many there are $ b $s with $ b\in  \mathbb{F}_{q} $ such that $ f(X)=- b^{p^{\alpha}}$ is solvable, which is answered by the following lemma:
  
  \begin{sec3_lemma3}[\cite{Bib7}, Lemma 21]\label{s3_lemma3}
  Set $ f(X)=X^{p^{2\alpha}} +X $ and 
  \begin{equation*}
  S=\lbrace b\in \mathbb{F}_{q}: f(X)=- b^{p^{\alpha}} \ \text{is solvable in }\   \mathbb{F}_{q} \rbrace.
  \end{equation*}
  If $ m/d $ is even, then $ | S |=p^{e-2d} $.
  \end{sec3_lemma3}
  
  For $ a\in \mathbb{F}_{p} $, the definiting set in \cite{Bib7} is given by
  \begin{equation}\label{defining_set_a}
  D_{a}=\biggl\lbrace x\in \mathbb{F}_{q}^{*} : Tr(x^{p^{\alpha}+1}) =a\biggr\rbrace.
  \end{equation}
  
  Let 
  \begin{equation}\label{codewords_length}
  n_{a}=
  \begin{cases}
  | D_{a} \cup \lbrace 0\rbrace | &\qquad\text{if}\ a=0,\\
  | D_{a}|,&\qquad\text{if}\ a\ne 0.
  \end{cases}
  \end{equation}
  
  It is clear that $ n_{a} $ of (\ref{codewords_length}) gives the length of codewords of the linear codes $ \mathcal{C}_{D_{a}} $ defined in (\ref{wang_LD}) and constructed by using the \textit{defining set} of (\ref{defining_set_a}), whose values were explicitly computed in \cite{Bib7}, resumed by the following lemma:
  
  \begin{sec3_lemma4}[\cite{Bib7}]\label{s3_lemma4}
  \begin{equation*}
  n_{0}=
  \begin{cases}
  p^{e-1}-(p-1)p^{m-1},&\qquad\text{if}\ m/d \ \text{is odd},\\
  p^{e-1}-(p-1)p^{m+d-1},&\qquad\text{if}\ m/d \ \text{is even}.\\
  \end{cases}
  \end{equation*}
  \begin{equation*}
    n_{a}=
    \begin{cases}
    p^{e-1}+p^{m-1},&\qquad\text{if}\ m/d \ \text{is odd},\\
    p^{e-1}+p^{m+d-1},&\qquad\text{if}\ m/d \ \text{is even}.\\
    \end{cases}
  \end{equation*}
  \end{sec3_lemma4}
  
  Suppose that $ f(X)=- b^{p^{\alpha}}$ is solvable for the case $ m/d $ is even, how many are there such $ b $s with $ b\in \mathbb{F}_{q} $ such that $ Tr(\mathfrak{S}_{b})=a $? (see Corollary \ref{s3_cor1}) This question is answered by the next lemma:
  
  \begin{sec3_lemma5}\label{s3_lemma5}
   Let $ q=p^{e} $,  $ e=2m $,  and suppose that $ m/d $ is even, where $ p $ is an odd prime and $ d=\gcd(\alpha,e) $. In addition, suppose that $ f(X)=X^{p^{2\alpha}} +X=- b^{p^{\alpha}} $ is solvable with nonempty solution set $ \mathfrak{S}_{b} $ for each $ b $ in the following set $  S $: 
    \begin{equation*}
    S=\biggl\lbrace b\in \mathbb{F}_{q}: Tr(\mathfrak{S}_{b})=a\biggr\rbrace,
    \end{equation*}
    where $ a\in \mathbb{F}_{p} $. Then $ | S |=n_{a}/p^{2d} $. 
  \end{sec3_lemma5}
  \begin{proof}
  By Corollary \ref{s3_cor1}, it is clear that the non empty solution set $ \mathfrak{S}_{b} $, of the equation $ f(X)=X^{p^{2\alpha}} +X=- b^{p^{\alpha}} $, is contained in $ D_{a} $ of (\ref{defining_set_a}). i.e., $ \mathfrak{S}_{b}\subseteq D_{a} $. By Theorem 4.1 in \cite{Bib9}, $ |\mathfrak{S}_{b}| =p^{2d} $, hence $ | \mathfrak{S}_{b}|$ divides $n_{a} $. For each $ \mathfrak{S}_{b}\subseteq D_{a} $ corresponds to an unique $ b $ such taht $ f(X)=- b^{p^{\alpha}}   $ is solvable and $ Tr(\mathfrak{S}_{b})=a $. The proof is complete.
  \end{proof}
  
  \begin{sec3_lemma6}\label{s3_lemma6}
  Let $ q=p^{e} $,  $ e=2m $,  and suppose that $ m/d $ is odd, where $ p $ is an odd prime and $ d=\gcd(\alpha,e) $. Then, $ f(X)=X^{p^{2\alpha}} +X $ is a permutation polynomial over $ \mathbb{F}_{q} $, and $ f(X)=X^{p^{2\alpha}} +X=- b^{p^{\alpha}} $ has an unique solution for each $ b\in  \mathbb{F}_{q}$. Denote its unique solution as $ x_{0} $ for the given $ b $. In addition, let 
  \begin{equation*}
  S=\biggl\lbrace b\in  \mathbb{F}_{q} : Tr(x_{0}^{p^{\alpha}+1})=a \biggr\rbrace,
  \end{equation*}
  with $ a\in  \mathbb{F}_{p} $. Then, $ |S|=n_{a} $.
  \end{sec3_lemma6}
  \begin{proof}
  Since $ x_{0} $ is unique for the given $ b $, when $ b $ ranges over $ \mathbb{F}_{q} $, $ x_{0} $ does too. Hence,
 $ S=\lbrace b\in  \mathbb{F}_{q} : Tr(x_{0}^{p^{\alpha}+1})=a \rbrace
 =\lbrace x\in  \mathbb{F}_{q} : Tr(x^{p^{\alpha}+1})=a \rbrace. $
  Therefore, $ |S|=|\lbrace x\in  \mathbb{F}_{q} : Tr(x^{p^{\alpha}+1})=a \rbrace|=n_{a} $.
  \end{proof}

  The next exponential sum over $ \mathbb{F}_{q} $ was defined in \cite{Bib7} in order to compute the Hamming weights of the code $ \mathcal{C}_{D_{a}} $ of (\ref{wang_LD}) with the \textit{defining set} $ D_{a} $ given by (\ref{defining_set_a}).
  \begin{equation}\label{Nac_set}
  N_{b}(a,c)=\biggl\lbrace x\in \mathbb{F}_{q}: Tr(x^{p^{\alpha}+1})=a\ \text{and}\ Tr(bx)=c\biggr\rbrace.
  \end{equation}
  Let $ wt(\mathbf{c}_{b}) $ denote the Hamming weight of the codeword $ \mathbf{c}_{b} (b\in \mathbb{F}_{q}^{*}) $ of the code $ \mathcal{C}_{D_{a}} $. It is easy to check out 
  \begin{equation}\label{wt_formula}
  wt(\mathbf{c}_{b})=n_{a}-|N_{b}(a,c)|.
  \end{equation}
  
  The following  lemmas give the explicit values of $ N_{b}(a,c) $ according to if $ m/d $ is odd or even.

   \begin{sec3_lemma12_a0}[\cite{Bib7}, Lemma 30]\label{Na0_md_even}
             Suppose that $ m/d $ is even, and $ a\in \mathbb{F}_{p}^{*}$.  For $ b\in \mathbb{F}_{q}^{*}  $, if 
             $  f(X)=X^{p^{2\alpha}} +X=- b^{p^{\alpha}} $ has no solution in $ \mathbb{F}_{q} $, then 
             \begin{equation*}
             |N_{b}(a,0)|=p^{e-2}+p^{m+d-2},
             \end{equation*}
             else
             \begin{equation*}
             |N_{b}(a,0)|=
             \begin{cases}
             p^{e-2}+p^{m+d-1},&\qquad\text{if}\ Tr\bigl(\gamma^{p^{\alpha}+1}\bigr)=0,\\
             p^{e-2}-p^{m+d-1}\widehat{\eta}\bigl(-aTr(\gamma^{p^{\alpha}+1})\bigr),&\qquad\text{if}\ Tr(\gamma^{p^{\alpha}+1}))\ne 0,
             \end{cases}
             \end{equation*}
             where $ \gamma $ is one of the  solutions of $ f(X)=X^{p^{2\alpha}} +X=- b^{p^{\alpha}} $, i.e., $ \gamma\in \mathfrak{S}_{b} $.
             \end{sec3_lemma12_a0}
  
   \begin{sec3_lemma12}[\cite{Bib7}, Lemma 31]\label{Nac_md_even}
          Suppose that $ m/d $ is even, and $ (a,c)\in \mathbb{F}_{p}^{*}\times\mathbb{F}_{p}^{*} $.  For $ b\in \mathbb{F}_{q}^{*}  $, if 
          $  f(X)=X^{p^{2\alpha}} +X=- b^{p^{\alpha}} $ has no solution in $ \mathbb{F}_{q} $, then 
          \begin{equation*}
          |N_{b}(a,c)|=p^{e-2}+p^{m+d-2},
          \end{equation*}
          else
          \begin{equation*}
          |N_{b}(a,c)|=
          \begin{cases}
          p^{e-2},&\qquad\text{if}\ Tr\bigl(\gamma^{p^{\alpha}+1}\bigr)=0,\\
          p^{e-2},&\qquad\text{if}\ Tr\bigl(\gamma^{p^{\alpha}+1}\bigr)=c^{2}/(4a),\\
          p^{e-2}-p^{m+d-1}\widehat{\eta}\bigl(c^{2}-4aTr(\gamma^{p^{\alpha}+1})\bigr),&\qquad\text{otherwise},
          \end{cases}
          \end{equation*}
          where $ \gamma $ is one of the  solutions of $ f(X)=X^{p^{2\alpha}} +X=- b^{p^{\alpha}} $, i.e., $ \gamma\in \mathfrak{S}_{b} $.
          \end{sec3_lemma12}

Now we are ready to prove \textbf{Theorem} \ref{s2_thm1}-\ref{s2_thm4}, but we only prove \textbf{Theorem} \ref{s2_thm4} since proofs for other three theorems are similar and omitted due to limited space.
\begin{proof}[Proof of \textbf{Theorem} \ref{s2_thm4}]
\begin{enumerate}[(I)]
\item Case for $ b\in \mathbb{F}_{q}^{*}, f(X)=X^{p^{2\alpha}} +X=- b^{p^{\alpha}} $ has no solution in $ \mathbb{F}_{q} $.

From Lemma \ref{s3_lemma3}, the number of $  b\in \mathbb{F}_{q}^{*} $ such that $ f(X)=X^{p^{2\alpha}} +X=- b^{p^{\alpha}} $ has no solution is equal to $ p^{e}-p^{e-2d} $. From Lemma \ref{Na0_md_even} and \ref{Nac_md_even} the parts for this case, for each symbol $ i\in \mathbb{F}_{p} $ (see Section 1), its exponent is $ p^{e-2}+p^{m+d-2} $, therefore \textit{the contribution of this case} to the complete weight enumerator is
\begin{equation}\label{proof_thm4_cwe01}
C_{1}=(p^{e}-p^{e-2d})\prod_{\substack{0\leq i\leq p-1}}w_{i}^{p^{e-2}+p^{m+d-2}}.
\end{equation}
From (\ref{proof_thm4_cwe01}), the corresponding Hamming weight, denoted by $ wt(C_{1}) $, and the multiplicity, denoted by $ A_{C_{1}} $, are respectively
\begin{equation}\label{proof_thm4_hamming01}
\begin{split}
wt(C_{1})&=(p-1) (p^{e-2}+p^{m+d-2}),\\
A_{C_{1}}&=p^{e}-p^{e-2d}.
\end{split}
\end{equation}
\item Case where for $ b\in \mathbb{F}_{q}^{*}, f(X)=X^{p^{2\alpha}} +X=- b^{p^{\alpha}} $ has a solution $ \gamma $ such that $ Tr\bigl(\gamma^{p^{\alpha}+1}\bigr)=0 $.

From Lemma \ref{s3_lemma4}  and Lemma \ref{s3_lemma5}, the number of $ b\in \mathbb{F}_{q}^{*}$ such that $f(X)=X^{p^{2\alpha}} +X=- b^{p^{\alpha}} $ has a solution $ \gamma $ and $ Tr\bigl(\gamma^{p^{\alpha}+1}\bigr)=0 $, is equal to $ p^{e-2d-1}-(p-1)p^{m-d-1}-1 $. By Lemma \ref{Na0_md_even}, the exponent of symbol $ c $ is $ p^{e-2}+p^{m+d-1} $. From Lemma \ref{Nac_md_even}, the exponents of other symbols other than $ c $ are $ p^{e-2} $. Hence, \textit{the contribution of this case} to the complete weight enumerator is
\begin{equation}\label{proof_thm4_cwe02}
C_{2}=\bigl(p^{e-2d-1}-(p-1)p^{m-d-1}-1\bigr)w_{c}^{p^{e-2}+p^{m+d-1}}\prod_{\substack{0\leq i\leq p-1\\i\ne c}}w_{i}^{p^{e-2}}.
\end{equation}
From (\ref{proof_thm4_cwe02}), the corresponding Hamming weight and the multiplicity are respectively
\begin{equation}\label{proof_thm4_hamming02}
\begin{split}
wt(C_{2})&=(p-1)p^{e-2}+p^{m+d-1},\\
A_{C_{2}}&=p^{e-2d-1}-(p-1)p^{m-d-1}-1.
\end{split}
\end{equation}
\item Case where for $ b\in \mathbb{F}_{q}^{*}, f(X)=X^{p^{2\alpha}} +X=- b^{p^{\alpha}} $ has a solution $ \gamma $ such that $ Tr\bigl(\gamma^{p^{\alpha}+1}\bigr)=\frac{c^{2}}{4a} $. Remark that $ \biggl(\dfrac{a}{p}\biggr)=\biggl(\dfrac{Tr\bigl(\gamma^{p^{\alpha}+1}\bigr)}{p}\biggr) $ since $ Tr\bigl(\gamma^{p^{\alpha}+1}\bigr)=\frac{c^{2}}{4a} $. From Lemma \ref{s3_lemma5} at first and Lemma \ref{s3_lemma4}, the number of $ b\in \mathbb{F}_{q}^{*}$ such that $f(X)=X^{p^{2\alpha}} +X=- b^{p^{\alpha}} $ has a solution $ \gamma $ and $ Tr\bigl(\gamma^{p^{\alpha}+1}\bigr)=\frac{c^{2}}{4a}$, is equal to $ p^{e-2d-1}+p^{m-d-1} $.
To ease the proof, define
\begin{equation}\label{proof_thm4_Nacnu}
N_{b}(a,c:\rho)=\biggl\lbrace x\in \mathbb{F}_{q}: Tr(x^{p^{\alpha}+1})=a\ \text{and}\ Tr(bx)+c=\rho\biggr\rbrace.
\end{equation}
It is clear that $ N_{b}(a,c:\rho)= N_{b}(a,\bar{c})$ where $ \bar{c}=\rho-c $. Regarding $ \bar{c} $, three cases may be distinguished: $ \bar{c}=0,\bar{c}^{2}=4aTr\bigl(\gamma^{p^{\alpha}+1}\bigr)=c^{2} $ and $ \bar{c}^{2}\ne 4aTr\bigl(\gamma^{p^{\alpha}+1}\bigr)=c^{2} $.
\begin{itemize}
\item $ \bar{c}=0 $, corresponding to the symbol $ c $ whose exponent can be determined by Lemma \ref{Na0_md_even} and equals to $ p^{e-2}-p^{m+d-1}\widehat{\eta}\bigl(-aTr(\gamma^{p^{\alpha}+1})\bigr)=p^{e-2}-p^{m+d-1}\biggl(\dfrac{-1}{p}\biggr) $.
\item $ \bar{c}=4aTr\bigl(\gamma^{p^{\alpha}+1}\bigr)=c^{2} $, corresponding to the symbol $ 0,2c\pmod p $ whose exponents can be determined by Lemma \ref{Nac_md_even} and equal to $ p^{e-2} $.
\item $ \bar{c}\ne 4aTr\bigl(\gamma^{p^{\alpha}+1}\bigr)=c^{2} $, corresponding to the symbol $ (\bar{c}+c)\pmod p (\bar{c}\ne 0,\pm c)$ whose exponents can be determined by Lemma \ref{Nac_md_even} and equal to 
\begin{equation*}
p^{e-2}-p^{m+d-1}\widehat{\eta}\bigl(\bar{c}^{2}-4aTr(\gamma^{p^{\alpha}+1})\bigr)=
p^{e-2}-p^{m+d-1}\biggl(\dfrac{\bar{c}^{2}-c^{2}}{p}\biggr).
\end{equation*}
\end{itemize}
\textit{The contribution of this case} to the complete weight enumerator is obtained by gathering the above results:
\begin{equation}\label{proof_thm4_cwe03}
C_{3}=(p^{e-2d-1}+p^{m-d-1})w_{c}^{p^{e-2}-\bigl(\frac{-1}{p}\bigr)p^{m+d-1}}\prod_{\substack{1\leq i\leq p-1}}w_{i+c}^{p^{e-2}-\bigl(\frac{i^{2}-c^{2}}{p}\bigr)p^{m+d-1}}.
\end{equation}
We use Corollary \ref{sec3_Legendre_symbol_Lpc} to compute the Hamming weight of (\ref{proof_thm4_cwe03}).
\begin{equation}\label{proof_thm4_hamming_weight_c3}
\begin{split}
wt(C_{3})&=\biggl(p^{e-2}-\biggl(\frac{-1}{p}\biggr)p^{m+d-1}\biggr)+\sum_{i=1}^{p-1}\biggl(p^{e-2}-\biggl(\frac{i^{2}-c^{2}}{p}\biggr)p^{m+d-1}\biggr)-p^{e-2}\\
&=(p-1)p^{e-2}-\biggl(\biggl(\frac{-1}{p}\biggr)+L_{p}(c)\biggr)p^{m+d-1}\\
&=(p-1)p^{e-2}+p^{m+d-1}.
\end{split}
\end{equation}
Note that the multiplicity of $ wt(C_{3}) $ is $ A_{C_{3}}=  p^{e-2d-1}+p^{m-d-1}$. 
\item Case where for $ b\in \mathbb{F}_{q}^{*}, f(X)=X^{p^{2\alpha}} +X=- b^{p^{\alpha}} $ has a solution $ \gamma $ such that $ Tr\bigl(\gamma^{p^{\alpha}+1}\bigr)\ne \frac{c^{2}}{4a} $. Let $ i= Tr\bigl(\gamma^{p^{\alpha}+1}\bigr)$, then $ 1\leq i\leq p-1 $ and $ i\ne \frac{c^{2}}{4a} $. Furthermore, suppose $ \biggl(\dfrac{i}{p}\biggr)=\biggl(\dfrac{a}{p}\biggr) $. From Lemma \ref{s3_lemma5} at first and Lemma \ref{s3_lemma4}, the number of $ b\in \mathbb{F}_{q}^{*}$ such that $f(X)=X^{p^{2\alpha}} +X=- b^{p^{\alpha}} $ has a solution $ \gamma $ and $ Tr\bigl(\gamma^{p^{\alpha}+1}\bigr)\ne \frac{c^{2}}{4a} $, is equal to $ p^{e-2d-1}+p^{m-d-1} $. Regarding $ \bar{c} $ defined in(\ref{proof_thm4_Nacnu}), three cases can be distinguished: $ \bar{c}=0,\bar{c}^{2}=4ai $ and $ \bar{c}^{2}\ne 4ai $.
\begin{itemize}
\item $ \bar{c}=0 $, corresponding to the symbol $ c $ whose exponent can be determined by Lemma \ref{Na0_md_even} and equals to $ p^{e-2}-p^{m+d-1}\widehat{\eta}\bigl(-aTr(\gamma^{p^{\alpha}+1}))\bigr)=p^{e-2}-p^{m+d-1}\biggl(\dfrac{-1}{p}\biggr) $.
\item $ \bar{c}=4ai $, corresponding to the symbol $ (c\pm 2\sqrt{ai})\pmod p $ whose exponents can be determined by Lemma \ref{Nac_md_even} and equal to $ p^{e-2} $. Remark that the square root of $ 4ai $ exists since $ \biggl(\dfrac{i}{p}\biggr)=\biggl(\dfrac{a}{p}\biggr) $.
\item $ \bar{c}\ne 4ai $, corresponding to the symbol $ (\bar{c}+c)\pmod p (\bar{c}\ne 0,\pm 2\sqrt{ai})$ whose exponents can be determined by Lemma \ref{Nac_md_even} and equal to 
\begin{equation*}
p^{e-2}-p^{m+d-1}\widehat{\eta}\bigl(\bar{c}^{2}-4aTr(\gamma^{p^{\alpha}+1})\bigr)=
p^{e-2}-p^{m+d-1}\biggl(\dfrac{\bar{c}^{2}-4ai}{p}\biggr).
\end{equation*}
\end{itemize}
According to above discussion, \textit{the contribution of this case} to the complete weight enumerator is
\begin{equation}\label{proof_thm4_cwe04}
C_{4}=\sum_{\substack{1\leq i\leq p-1\\i\ne (4a)^{-1}c^{2}\\\bigl(\frac{i}{p}\bigr)=\bigl(\frac{a}{p}\bigr)}}(p^{e-2d-1}+p^{m-d-1})w_{c}^{p^{e-2}-\bigl(\frac{-1}{p}\bigr)p^{m+d-1}}\prod_{\substack{1\leq j\leq p-1}}w_{j+c}^{p^{e-2}-\bigl(\frac{j^{2}-4ai}{p}\bigr)p^{m+d-1}}.
\end{equation}
We use Corollary \ref{sec3_Legendre_symbol_spai} to compute the Hamming weight of (\ref{proof_thm4_cwe04}):
\begin{equation*}
\begin{split}
wt(C_{4})&=\biggl(p^{e-2}-\biggl(\frac{-1}{p}\biggr)p^{m+d-1}\biggr)+\sum_{j=1}^{p-1}\biggl(p^{e-2}-\biggl(\frac{j^{2}-4ai}{p}\biggr)p^{m+d-1}\biggr)-\biggl(p^{e-2}-\biggl(\frac{c^{2}-4ai}{p}\biggr)p^{m+d-1}\biggr)\\
&=(p-1)p^{e-2}+\biggl(\biggl(\frac{c^{2}-4ai}{p}\biggr)-\biggl(\frac{-1}{p}\biggr)-S_{p}(a,i)\biggr)p^{m+d-1}\\
&=(p-1)p^{e-2}+\biggl(1+\biggl(\frac{c^{2}-4ai}{p}\biggr)\biggr)p^{m+d-1}.
\end{split}
\end{equation*}
Hence, we thus obtain
\begin{equation}\label{proof_thm4_hamming_weight_c4}
wt(C_{4})=
\begin{cases}
(p-1)p^{e-2},&\qquad\text{if}\ \biggl(\frac{c^{2}-4ai}{p}\biggr)=-1,\\
(p-1)p^{e-2}+2p^{m+d-1},&\qquad\text{if}\ \biggl(\frac{c^{2}-4ai}{p}\biggr)=1.
\end{cases}
\end{equation}
From Lemma \ref{sec3_npac}, the multiplicity of (\ref{proof_thm4_hamming_weight_c4}) is:
\begin{equation}\label{proof_thm4_wtmultiplicity_c4}
A_{C_{4}}=
\begin{cases}
\#N_{p}^{-}(a,c) (p^{e-2d-1}+p^{m-d-1}),&\qquad\text{if}\ \biggl(\frac{c^{2}-4ai}{p}\biggr)=-1,\\
\#N_{p}^{+}(a,c) (p^{e-2d-1}+p^{m-d-1}),&\qquad\text{if}\ \biggl(\frac{c^{2}-4ai}{p}\biggr)=1.
\end{cases}
\end{equation}
\item Case where for $ b\in \mathbb{F}_{q}^{*}, f(X)=X^{p^{2\alpha}} +X=- b^{p^{\alpha}} $ has a solution $ \gamma $ such that $ Tr\bigl(\gamma^{p^{\alpha}+1}\bigr)\ne \frac{c^{2}}{4a} $. Let $ i= Tr\bigl(\gamma^{p^{\alpha}+1}\bigr)$, then $ 1\leq i\leq p-1 $ and $ i\ne \frac{c^{2}}{4a} $. Furthermore, suppose $ \biggl(\frac{i}{p}\biggr)\ne\biggl(\frac{a}{p}\biggr) $. The analysis of the actual case is similar to case (IV), we omit the detail and write down the result:
\begin{equation}\label{proof_thm4_cwe05}
C_{5}=\sum_{\substack{1\leq i\leq p-1\\\bigl(\frac{i}{p}\bigr)\ne \bigl(\frac{a}{p}\bigr)}}(p^{e-2d-1}+p^{m-d-1}) w_{c}^{p^{e-2}+(\frac{-1}{p})p^{m+d-1}}
        \prod_{\substack{1\leq j\leq p-1}}w_{j+c}^{p^{e-2}-\bigl(\frac{j^{2}-4ai}{p}\bigr)p^{m+d-1}}
\end{equation}
The Hamming weight of $ C_{5} $ is 
\begin{equation}\label{proof_thm4_hamming_weight_c5}
wt(C_{5})=
\begin{cases}
(p-1)p^{e-2},&\qquad\text{if}\ \biggl(\frac{c^{2}-4ai}{p}\biggr)=-1,\\
(p-1)p^{e-2}+2p^{m+d-1},&\qquad\text{if}\ \biggl(\frac{c^{2}-4ai}{p}\biggr)=1.
\end{cases}
\end{equation}
From Lemma \ref{sec3_npac}, the multiplicity of (\ref{proof_thm4_hamming_weight_c5}) is:
\begin{equation}\label{proof_thm4_wtmultiplicity_c5}
A_{C_{5}}=
\begin{cases}
\#M_{p}^{-}(a,c) (p^{e-2d-1}+p^{m-d-1}),&\qquad\text{if}\ \biggl(\frac{c^{2}-4ai}{p}\biggr)=-1,\\
\#M_{p}^{+}(a,c) (p^{e-2d-1}+p^{m-d-1}),&\qquad\text{if}\ \biggl(\frac{c^{2}-4ai}{p}\biggr)=1.
\end{cases}
\end{equation}

Finally, in (\ref{wang_LD}) set $ x=0 $, we obtain the singleton codeword of which each component is equal to the symbol $ c $. By Lemma \ref{s3_lemma4}, the exponent of the symbol $ c $ is $p^{e-1}+p^{m+d-1}  $, so \textit{the contribution of this case to the complete weight enumerator} is 
\begin{equation}\label{proof_thm4_cwe0}
C_{0}=w_{c}^{p^{e-1}+p^{m+d-1}}.
\end{equation}
It is obvious that $ wt(C_{0})=p^{e-1}+p^{m+d-1} $, and the multiplicity  $ A_{C_{0}}=  1 $.
\end{enumerate}
By (\ref{proof_thm4_cwe01}), (\ref{proof_thm4_cwe02}), (\ref{proof_thm4_cwe03}), (\ref{proof_thm4_cwe04}),  (\ref{proof_thm4_cwe05}), and (\ref{proof_thm4_cwe0}), the complete weight enumerator of the code given by \textbf{Theorem} \ref{s2_thm4} is
\begin{equation*}
\mathit{CWE}(\mathcal{C}_{D})=C_{0}+C_{1}+C_{2}+C_{3}+C_{4}+C_{5},
\end{equation*}
which is identical to (\ref{Fac5Even}). By gathering the results from (\ref{proof_thm4_hamming01}), (\ref{proof_thm4_hamming02}), (\ref{proof_thm4_hamming_weight_c3}), (\ref{proof_thm4_hamming_weight_c4}) associated with (\ref{proof_thm4_wtmultiplicity_c4}), and (\ref{proof_thm4_hamming_weight_c5}) associated with (\ref{proof_thm4_wtmultiplicity_c5}), the Hamming weight and its multiplicity of each contribution to the complete weight enumerator, $ C_{0}-C_{5} $, are listed in Table \ref{tab_sec3_proof_thm4} for $ p\equiv 1\pmod 4 $. Remark that for $ p\equiv 3\pmod 4 $, it leads to a similar table. Table \ref{tab_s2_thm4} that gives the weight distribution of Theorem \ref{s2_thm4} can be derived directly from Table \ref{tab_sec3_proof_thm4} and the one for  $ p\equiv 3\pmod 4 $ (here omitted). The proof is completed.
\begin{table}[tbh]
       \caption{Complete weight enumerators and their weight distributions of  Theorem \ref{s2_thm4}}
              \label{tab_sec3_proof_thm4}
              \begin{center}
              \begin{tabular}{|c|c|c|}
              \hline
               $ C_{i} $ & weight $ wt(C_{i}) $  & multiplicity $ A_{C_{i}} $ \\
               \hline
               $ C_{0} $ & $ p^{e-1}+p^{m+d-1} $ & $ 1 $\\
               \hline
               $ C_{1}  $ & $ (p-1) \bigl(p^{e-2}+p^{m+d-2}\bigr) $ & $ p^{e}-p^{e-2d} $\\
               \hline
                $ C_{2} $ & $ (p-1)p^{e-2}+p^{m+d-1} $ & $ p^{e-2d-1}-(p-1)p^{m-d-1}-1 $\\
              \hline
               $ C_{3} $ & $ (p-1)p^{e-2}+p^{m+d-1} $ & $ p^{e-2d-1}+p^{m-d-1} $\\
              \hline
               $ C_{4} $ & $ (p-1)p^{e-2} $ & $ \frac{p-1}{4}\bigl(p^{e-2d-1}+p^{m-d-1}\bigr) $\\
               \hline
               $ C_{4} $ & $ (p-1)p^{e-2}+2p^{m+d-1} $ & $ \frac{p-5}{4}\bigl(p^{e-2d-1}+p^{m-d-1}\bigr) $\\
               \hline
               $ C_{5} $ & $ (p-1)p^{e-2} $ & $\frac{p-1}{4}\bigl(p^{e-2d-1}+p^{m-d-1}\bigr)  $\\
               \hline
               $ C_{5} $ & $ (p-1)p^{e-2}+2p^{m+d-1} $ & $ \frac{p-1}{4}\bigl(p^{e-2d-1}+p^{m-d-1}\bigr) $\\
               \hline
              \end{tabular}
              \end{center}
        \end{table}
\end{proof}

\end{document}